\documentclass[12pt]{article}
\usepackage{amssymb,amsmath,epsfig}
\usepackage{cite}
 \allowdisplaybreaks

\begin{document}
\title{\bf Influence of Electromagnetic Field on Hyperbolically Symmetric Source}
\author{M. Z. Bhatti \thanks{mzaeem.math@pu.edu.pk}, Z. Yousaf \thanks{zeeshan.math@pu.edu.pk} and Z. Tariq
\thanks{zohatariq24@yahoo.com}\\
Department of Mathematics, University of the Punjab,\\
Quaid-i-Azam Campus, Lahore-54590, Pakistan.}
\date{}
\maketitle
\begin{abstract}
We pursue a coherent analysis of hyperbolically symmetric static sources by extending the work of Herrera \cite{26} to the case of electromagnetic field. We deeply analyze the impact of such a force on the physical characteristics of the hyperbolically symmetric spacetime under consideration. Setting off the Einstein's gravitational equations, we particularize the stress-energy tensor by keeping in mind the constituents of tetrad field in Minkowski coordinate system. Hyperbolically symmetric source has a vacuole in the vicinity of its center, i.e., such a distribution
is unable to fill the central zone of symmetry. In all the stellar expressions, the energy density of the source comes out to be negative which demonstrates
that quantum effects must be involved along with certain extreme restrictions to explicate any physical application of such a hyperbolically symmetric source.
Various explicit exact solutions along with their corresponding generating functions are also worked out.
\end{abstract}
{\bf Keywords:} Electromagnetic field, Non-spherical sources, Relativistic fluids.\\
{\bf PACS:} 41.20; 04.40.Nr; 04.40.-b.

\section{Introduction}

Albert Einstein finalized his theory of general relativity (GR) in 1915, and since then, its impact has been profoundly affecting the research in physics,
mathematics and cosmology. He theorized that the matter is capable of
influencing the space and conjectured that if the Sun dies out suddenly, the indication for the planets to terminate orbiting it would rationally take some time. He proposed the notion that space and time are intricately linked and the events that happen at a time to a viewer could happen at different time for another
viewer. The basis of Einstein's GR is laid upon the spacetime structure, the local casuality, the equivalence principle and the local coordinate frames. It sheds light on the fact that spacetime is a differentiable manifold having four dimensions and Lorentzian signature $(+,-,-,-,-)$ or $(-,+,+,+)$. Einstein coupled the matter sources (represented by the stress-energy tensor) and the metric via his non-linear equations of motion known famously as the Einstein's field equations. His GR functions correctly and resolves previously indicated issues regarding the space and time.

Hyperbolically symmetric spacetimes tend to provide a feasible comprehension of unsolved astronomical perplexities. Harrison \cite{1} yielded few exact solutions of gravitational equations by utilizing the interchanging variable approach for hyperbolic symmetry. Using variable separation method, several solutions to Geld equations were also obtained by keeping in view the vacuum space. Feasible geometric interpretation was also provided to comprehend the solutions that were thirty in number. Madler \cite{2} considered a null hypersurface and obtained metric functions along with their first order derivatives as the initial data. Further, the gravitational equations are integrated and a set of equations is solved on the boundary, i.e., the null hypersurface. As an example, the double null Israel black hole (BH) solution is expressed, particularly for the case of spherically symmetric system and vacuum.
Gaudin et al. \cite{3} investigated ceratin static solutions of gravitational equations for a massless scalar field by connecting solutions with
Kantowski-Sachs solutions. They also explored the characteristics of vacuum hyperbolic spacetime in detail. Lobo and Mimiso \cite{4} performed an analysis
based on possibility of existence of tunnels in a hyperbolically symmetric spacetime. By the addition of exotic matter in vacuum solutions, they attained
solutions for pseudo-spherically symmetric static spacetime. They also demonstrated physical attributes of such solutions via mathematical embedding. Ren \cite{5}
worked out few analytic solutions by taking into consideration the hyperbolic BHs with no planar or spherical counterparts. Few applications of such hyperbolic
BHs were also illustrated. Furthermore, a C-metric solution is also provided that generalizes the hyperbolic BHs having scalar hair.

The occurrence of vacuoles or cavities in the spherically symmetric configurations was first explored by Skripkin \cite{6} while studying their evolutionary stages. Such cavities play a vital role to model voids in space and have been discussed previously by many researchers in detail. Herrera et al. \cite{6*} examined
spherically symmetric fluids and studied the evolution of cavities present in such relativistic sources. By assuming that the
proper radial distance between two adjacent elements of the source has constant value during the evolutionary phase, the authors demonstrated few solutions and
contemplated a general formalism. Torres \cite{7} depicted few outcomes regarding voids and bubbles in universe. They obtained general consequences by considering
spherically symmetric voids and bubbles that possess null radiation, electric charge or cosmological constant. They inferred few restrictions under which such
bubbles or voids are generated via phase transition and obtained the conditions for radial pressures. They also reported some models with special focus on
inflationary ones. Bonnor and Chamorro \cite{8} modeled a void in expanding cosmos using spherical Minkowksi region for the case of Tolman dust metric. They
found few interesting results and presented one expanding void model and few non-expanding void models. Yousaf and Bhatti \cite{9} found few constraints regarding
dynamical instability of cylindrically symmetric system in the context of $f(R,T)$ gravity. They acquired the modified gravitational equations along with few other
stellar equations and discussed the part played by the expansion scalar. They concluded that the dark source terms appearing in the gravity model affect the
instability of the configuration. Herrera et al. \cite{10} investigated the dynamical instability of a relativistic anisotropic fluid with spherical geometry collapsing adiabatically under the constraint of zero expansion scalar. They performed an in-depth analysis of Newtonian as well as post Newtonian regimes and concluded that in both of these approximations, the adiabatic index that corresponds to the stiffness of the fluid is of no use.

Static solutions of the Einstein's gravitational equations must be acquired in order to explicate the evolution of gravitational configuration.
Zubairi et al. \cite{11} considered the compact objects and presented two solutions of gravitational equations as neutron stars and quarks. Taking into account the spherical distribution of source, they evaluated the solution of modified equations by taking finite values of cosmological constant. They also explored physical attributes such as masses and radii of deformed stellar structures and pointed out the differences between such systems and the standard spherical models. Berger \cite{12} evaluated the gravitational equations for the case of perfect fluid having spherical symmetry. The obtained
analytical solution was found to be dependent on the radial coordinate. As an example, the interior and exterior Schwarzschild solutions are re-acquired. Trenda and Fulling
\cite{13} found static vacuum solutions of Einstein's equations for a configuration with cylindrical symmetry. They matched the acquired vacuum solutions
with the interior solutions by considering that the interior solutions have non-zero pressure and density. Along with the string solution, they also found
few numerical solutions. Leibovitz \cite{14} obtained a formal solution of Einstein's equations for static spacetime incorporating arbitrary function dependent
on radial coordinate. They demonstrated a mapping from Newtonian solutions and considered infinite value of stress at the central region of the source.

From the orthogonal division of Riemann tensor, few scalar parameters appear that assist to figure out the characteristics of fluid source. Such scalar functions are named as Structure Scalars (SS) and have been utilized extensively in literature. Yousaf analyzed relativistic systems having spherical symmetry and
explored the role of $f(G,T)$ \cite{15} and Palatini $f(R)$ \cite{15a} modification of gravitational force on such systems. By manipulating the Misner-Sharp mass function along with the tidal forces and the
structure scalars, they investigated the evolutionary stages of radiating spheres. The part played by the structure scalars is examined in the absence and
presence of modified gravity terms. Few authors constructed modified scalar functions in modified to evaluate kinematical
variables and structure and evolution equations \cite{16,16a,16b,16c,16d,16e}. They divided the Riemann tensor orthogonally by utilizing Herrera's formalism \cite{29}. From such a division,
modified structure scalars are acquired which may be used to define the physical attributes of the source. Bhatti et al. \cite{18} investigated spherically
symmetric configurations by linking the fundamental properties of matter with the scalar functions (structure scalars) acquired from the breakdown of Riemann
tensor. They also attained few anisotropic static spheres to comprehend the static configurations in the context of metric $f(R)$ gravity. Bhatti and
Tariq \cite{19} studied spherically symmetric astronomical objects endowed with heat dissipation in the presence of electromagnetic field within the background of
GR. From the division of Riemann tensor, they procured five scalar parameters which are later found to be directly connected with the fluid characteristics like
energy density inhomogeneity, pressure anisotropy etc. Herrera et al. \cite{20} applied $1+3$ approach to present a set of stellar equations that control the
evolutionary phases and structure formation of cylindrically symmetric fluids having stress anisotropy. Few variables (e.g., dissipative flux and active
gravitational mass) play a key role in the dynamical inspection of such fluid sources.

A process in which the gas can neither exchange heat from its surroundings nor extract heat from any kind of internal source is termed as adiabatic process. The adiabatic index plays a key role
in understanding the resistance shown by the gas when some kind of force tries to compress it. Much work on the significance of adiabatic index can be found
in literature. Esculpi et al. \cite{21} acquired a class of static solutions of Einstein's gravitational equations for a static spherical geometry having homogeneous
energy density. They found that the solutions thus obtained were dependent upon two parameters connected with the number of degrees of fluid anisotropy.
They also investigated the stability of the acquired solutions via slow adiabatic contraction. Tooper \cite{22} dealt with hot massive non-rotating stars having
adiabatic temperature gradients by assuming that such stars are composed of mixture of ideal gas and radiation forming solution with two parameter family.
They concluded that for all the presented models with constant value of rest mass, the dynamical stability appears at the first relative maxima of the binding energy. Herrera et al. \cite{23} analyzed adiabatic contraction of spherical geometry with stress anisotropy and different degrees of anisotropy.
Making use of different initial conditions, the equation of motion is integrated. Few models are found to be more stable than the isotropic ones.
Ivanov \cite{24} expressed all the attributes of self-gravitating spheres fulfilling a second-order differential equation. A variety of constraints
are employed to procure novel solutions and to derive classical outcomes particularly, for dust and perfect fluids.

This article is the extension of the analysis carried out by Herrera et al. \cite{26} to the charged case. We evaluated the same stellar equations but in the presence of electromagnetic force to reveal the influence of such a force on the hyperbolically symmetric self-gravitating body. We have arranged our article in the following fashion. Section \textbf{2} incorporates the hyperbolically symmetric line element along with the description of matter
content and a general formalism needed to define a static anisotropic self-gravitating structure. Section \textbf{3} entails the Einstein's equations of motion
endowed with the electric charge, the definition of charged Misner-sharp mass and the equation of hydrostatic equilibrium. Section \textbf{4} employs the curvature
tensors and the mathematical expression for Tolman mass (the active gravitational mass) of the structure. Section \textbf{5} deals with the division of
Riemann tensor into three constituents which are later used to extract four scalar parameters. Section \textbf{6} comprises several static hyperbolically
symmetric explicit solutions along with their generating functions. Section \textbf{7} winds up the discourse with concluding statements.

\section{The Metric, The Source and Physical Variables}

A static hyperbolically symmetric source incorporating stress anisotropy is contemplated. Along with the assumption of its boundedness from exterior via surface $\Sigma^e$ defined as $r=r_{\Sigma^e}$, it is supposed that the matter is incapable of occupying the middle region of the geometry. For this reason, the middle region is indicated by a vacuole which in turn points towards the fact that the matter source is bounded also from the interior, defined mathematically as $r=r_{\Sigma^i}$. The general hyperbolically symmetric metric in polar coordinates is given as
\begin{equation}\label{1f}
ds^2=e^{\nu(r)}dt^2-e^{\lambda(r)}dr^2-r^2d\theta^2-r^2 sinh^2 \theta d\phi^2.
\end{equation}
We may resort to the constituents of electromagnetic stress-energy tensor to account for the electric charge around the astronomical configuration. Mathematically, it can be demonstrated as
\begin{equation}\nonumber
\tilde{S}_{\alpha\xi}= \frac{1}{4\pi} \left( -F^\delta_\alpha
F_{\xi\delta}+\frac{1}{4}
F^{\delta\sigma}F_{\delta\sigma}g_{\alpha\xi}\right),
\end{equation}
with $F_{\xi\delta}$, having the mathematical expression, $F_{\xi\delta}=\varphi_{\delta,\xi}-\varphi_{\xi,\delta}$ signifies the electromagnetic field tensor while $\varphi^\xi$, having value $\varphi^\xi=\varphi(r)\delta^\xi_0$, represents the four-potential. The four-current density, i.e., $J^\xi$ has the expression $J^\xi=\sigma(r) u^\xi$. The Greek letters $\varphi$ and $\sigma$ utilized in above expressions depict the scalar potential and charge density, respectively. The differential equations (also called Einstein-Maxwell equations) read
\begin{equation}\nonumber
F^{\alpha\xi}_{~~;\xi}=\mu_0 J^\alpha; \quad \quad \quad
F_{[{\alpha\xi;\nu}]}=0,
\end{equation}
where the magnetic permeability is represented by the symbol $\mu_0$. Following second-order differential equation is yielded
\begin{align}\nonumber
&\frac{\partial^2\varphi}{\partial r^2}+
\frac{\partial\varphi}{\partial r}\left[-
\frac{{\lambda}'}{2}-\frac{{\nu}'}{2}+
\frac{2}{r}\right]=\mu_0 \sigma
e^{\lambda+\frac{\nu}{2}}.
\end{align}
Differentiating w.r.t. $r$, we acquire
\begin{equation}\nonumber
\frac{\partial\varphi}{\partial r}=\frac{\tilde{s}
e^{\frac{\nu+\lambda}{2}}}{r^2},\quad
\textmd{where} \quad
\tilde{s}=\int_0^r \mu_0 \sigma r^2 e^{\lambda/2}
dr.
\end{equation}
The constituents of electromagnetic stress-energy tensor take the following form
\begin{align}\nonumber
\tilde{S}_{00}=\frac{\tilde{s}^2 e^\nu}{8\pi r^4};\quad \tilde{S}_{11}= - \frac{\tilde{s}^2
e^\lambda}{8\pi r^4};\quad  \tilde{S}_{22}=\frac{\tilde{s}^2}{8\pi r^2};\quad
\tilde{S}_{33}=\tilde{S}_{22}Sinh^2\theta.
\end{align}
To get an idea about the physical state of the matter source, we define stress-energy tensor $T_{\alpha\xi}$ as a combination of energy density $\mu$,
isotropic stress $P$  and anisotropic tensor $\Pi_{\alpha\xi}$ as
\begin{equation}\nonumber
T_{\alpha\xi}=(\mu+P)V_\alpha V_\xi-Pg_{\alpha\xi}+\Pi_{\alpha\xi},
\end{equation}
where $V_\xi=(e^{\nu/2},0,0,0)$ indicates the four-velocity of the matter in terms of comoving coordinates. In order to particularize the direction of axes
of Minkowski coordinate system, a group of four unit vectors (orthogonal to each other) termed as the orthonormal tetrad [see \cite{26} for details] is given
as
\begin{align}\nonumber
e_\eta^{(0)}=V_\eta; \quad e_\eta^{(1)}&=K_\eta=(0,-e^{\lambda/2},0,0); \quad e_\eta^{(2)}=L_\eta=(0,0,-r,0); \\\nonumber
 e_\eta^{(3)}&=S_\eta=(0,0,0,-r sinh\theta).
\end{align}
Utilizing the notion of Bondi \cite{27}, we write down coordinates representing a locally Minkowski frame as provided below
\begin{equation}\nonumber
d\tilde{\tau}=e^{\nu/2}dt; \quad d\tilde{x}=e^{\lambda/2}dr; \quad d\tilde{y}=rd\theta; \quad d\tilde{z}=r sinh\theta d\phi.
\end{equation}
The constituents of stress-energy tensor in covariant notation indicating the matter configuration are given as under
$$ \check{T}_{\alpha\xi}=
\begin{pmatrix}
\mu & 0 & 0 & 0\\
0 & P_{xx} & P_{xy} & 0\\
0 & P_{yx} & P_y{y} & 0\\
0 & 0 & 0 & P_{zz} \\
\end{pmatrix},
$$
The symbols $P_{xx},~P_{xy},~P_{yy},~P_{zz}$ indicate the stresses in the direction specified by the subscripts. This mathematical expression
for $\check{T}_{\alpha\xi}$ is used for the case of axial symmetry. Since, we are working with the hyperbolically symmetric case, it is noted that the stress along $xy$ direction, i.e., $P_{xy}$ disappears and $P_{xx}\neq P_{yy}=P_{zz}$.
Utilizing the Minkowski coordinates, the constituents of tetrad field take the following form
\begin{align}\nonumber
\check{e}_\eta^{(0)}&=\check{V}_\eta=(1,0,0,0); \quad \check{e}_\eta^{(1)}=\check{K}_\eta=(0,-1,0,0); \\\nonumber
\quad \check{e}_\eta^{(2)}&=\check{L}_\eta=(0,0,-1,0); \quad \check{e}_\eta^{(3)}=\check{S}_\eta=(0,0,0,-1),
\end{align}
by making use of which, we can write
\begin{equation}\nonumber
\check{T}_{\alpha\xi}=(\mu+P_{zz})\check{V}_\alpha \check{V}_\xi-P_{zz}\eta_{\alpha\xi}+(P_{xx}-P_{zz})\check{e}_\alpha^{(1)}\check{e}_\xi^{(1)}.
\end{equation}
Here, $\eta_{\alpha\xi}$ symbolize the Minkowski metric. Getting back to the coordinate components used in metric (\ref{1f}), the constituents of
stress-energy tensor using the physical parameters for the local Minkowski frame can be written as
\begin{equation}\nonumber
T_{\alpha\xi}=(\mu+P_{zz})V_\alpha V_\xi-P_{zz}g_{\alpha\xi}+(P_{xx}-P_{zz})e_\alpha^{(1)}e_\xi^{(1)}.
\end{equation}
The anisotropic tensor expressed as $\Pi_{\alpha\xi}$ can be specified as
\begin{equation}\nonumber
\Pi_{\alpha\xi}=\Pi\left(e_\alpha^{(1)}e_\xi^{(1)}+\frac{h_{\alpha\xi}}{3}\right),
\end{equation}
with $\Pi=P_{xx}-P_{zz}$ whereas the isotropic stress may be defined as $P=\frac{P_{xx}+2P_{zz}}{3}$.
To analyze the boundedness of the matter source from the exterior, the smooth matching of metric (\ref{1f}) with the following hyperbolically symmetric version of Reissner-Nordstr\"{o}m metric is performed
\begin{equation}\nonumber
ds^2=\left(\frac{2M}{R}-1+\frac{Q^2}{r^2}\right)dt^2-\left(\frac{2M}{R}-1+\frac{Q^2}{r^2}\right)^{-1}dR^2-R^2 d\theta^2-R^2 sinh^2\theta d\phi^2.
\end{equation}
For such a smooth matching, we utilize the Darmois conditions \cite{25} to acquire following set of conditions
\begin{align}\nonumber
e^{\nu_{\Sigma^e}}&=\frac{2M}{r_{\Sigma^e}}-1+\left(\frac{Q^2}{r^2}\right)_{\Sigma^e};\;\;\; e^{\lambda_{\Sigma^e}}=\frac{1}{\frac{2M}{r_{\Sigma^e}}-1+\left(\frac{Q^2}{r^2}\right)_{\Sigma^e}},\\\nonumber
P_{xx}(r_{\Sigma^e})&=0,\;\;\; \left( Q^2=\tilde{s}^2\right)_{\Sigma^e}.
\end{align}
Since the middle region is assumed to be covered by vacuole, the implementation of the Darmois condition yields
\begin{equation}\nonumber
e^{\nu_{\Sigma^i}}=1;\;\;\; e^{\lambda_{\Sigma^i}}=1;\;\;\; P_{xx}(r_{\Sigma^i})=0;\;\;\; m(r_{\Sigma^i})=0.
\end{equation}

\section{Einstein-Maxwell Field Equations}

The Einstein's equations of motion for the metric (\ref{1f}) in the presence of electromagnetic field are mentioned below
\begin{align}\label{2f}
8\pi\left(\mu+\frac{\tilde{s}}{8\pi r^4}\right)&=-\frac{1+e^{-\lambda}}{r^2}+\frac{\lambda'e^{-\lambda}}{r},\\\label{3f}
8\pi\left(P_r-\frac{\tilde{s}}{8\pi r^4}\right)&=\frac{1+e^{-\lambda}}{r^2}+\frac{\nu'e^{-\lambda}}{r},\\\label{4f}
8\pi\left(P_\bot+\frac{\tilde{s}}{8\pi r^4}\right)&=\frac{e^{-\lambda}}{2}\left(\nu''-\frac{\lambda'\nu'}{2}+\frac{\nu'^2}{2}
-\frac{\lambda'}{r}+\frac{\nu'}{r}\right).
\end{align}
We make use of the notations $P_{xx}=P_r$ and $P_{yy}=P_{zz}=P_\bot$.
An equation that indicates a balance between the internally acting pressure-gradient force and the externally acting force of gravity is denominated as
hydrostatic equilibrium equation. Its value under our considerations takes the following form
\begin{equation}\label{5f}
P_r'+\frac{\nu'}{2}(\mu+P_r)+\frac{2\Pi}{r}-\frac{\tilde{s}\tilde{s}'}{4\pi r^4}=0.
\end{equation}
For charged hyperbolically symmetric configurations, the mass function (predefined without charge in \cite{26}) takes the following form
\begin{equation}\label{6f}
m(r)=\frac{r}{2}\left(1+e^{-\lambda}-\frac{\tilde{s}^2}{r^2}\right).
\end{equation}
Substituting Eq. (\ref{6f}) into (\ref{2f}), we render
\begin{equation}\label{7f}
m(r)=-4\pi \int_0^r\left(\mu r^2+\frac{\tilde{s}\tilde{s}'}{4\pi r}\right)dr.
\end{equation}
From Eq. (\ref{6f}), it can be noticed that $m$ is necessarily a positive quantity. Bearing this in mind, it turns out from Eq. (\ref{7f}) that the
energy density $\mu$ should be negative which is the clear-cut violation of weak energy condition. Some useful remarks on the physical significance
of this outcome are mentioned in \cite{26}. Thus, Eq. (\ref{7f}) can be re-arranged as
\begin{equation}\nonumber
m(r)=4\pi\int_{r_{min}}^r\left( |\mu|r^2 -\frac{\tilde{s}\tilde{s}'}{4\pi r}\right)dr,
\end{equation}
where, we have substituted $\mu$ by $-|\mu|$ because of the fact that the energy density of matter source under consideration is negative. Making use
of Eqs. (\ref{3f}) and (\ref{6f}), we acquire
\begin{equation}\label{7*f}
\nu'=2\left\{\frac{4\pi P_r r^3-m-\frac{\tilde{s}^2}{r}}{r(2m-r+\frac{\tilde{s}^2}{r})}\right\}.
\end{equation}
Substituting it in Eq. (\ref{5f}), we attain the hydrostatic equilibrium condition in terms of fluid variables $|\mu|,~P_r,~\Pi$ and mass function
$m$ along with a factor indicating the influence of charge as follows
\begin{equation}\label{7**f}
P_r'+\left(\frac{4\pi P_r r^3-m-\frac{\tilde{s}^2}{r}}{r(2m-r+\frac{\tilde{s}^2}{r})}\right)(P_r-|\mu|)+\frac{2\Pi}{r}
-\frac{\tilde{s}\tilde{s}'}{4\pi r^4}=0
\end{equation}
Some physical consequences of this equation are discussed with detail in \cite{26}.

\section{The Curvature tensors and the Active Gravitational Mass}

The intrinsic measure of curvature of a manifold, i.e., a Riemann tensor can be expressed as
\begin{equation}\nonumber
R^\nu_{\eta\rho\mu}=C^\nu_{\eta\rho\mu}+\frac{1}{2}R^\nu_\rho
g_{\eta\mu}+\frac{1}{2}R_{\eta\rho}\delta^\nu_\mu+\frac{1}{2}R_{\eta\mu}\delta^\nu_\rho-\frac{1}{2}R^\nu_\mu
g_{\eta\rho}-\frac{1}{6}R\left(\delta^\nu_\rho
g_{\eta\mu}-g_{\eta\rho}\delta^\nu_\mu\right),
\end{equation}
where the Weyl tensor, the description of which is possible from its electric part only, is written as
\begin{equation}\nonumber
C_{\xi\nu\pi\lambda}=E^{\beta\delta}V^\rho V^\gamma
(g_{\xi\nu\rho\beta}
g_{\pi\lambda\gamma\delta}-\eta_{\xi\nu\rho\beta}
\eta_{\pi\lambda\gamma\delta}),
\end{equation}
with
\begin{equation}\nonumber
g_{\xi\nu\rho\beta}=g_{\xi\rho}g_{\nu\beta}-g_{\xi\beta}g_{\nu\rho},
\end{equation}
where $\eta_{\pi\lambda\gamma\delta}$ specifies the Levi-Civita tensor. Making use of the tetrad, we can express $E_{\alpha\xi}$ as
\begin{equation}\nonumber
E_{\alpha\xi}=\varepsilon\left(e_\alpha^{(1)}e_\xi^{(1)}+\frac{1}{3} h_{\alpha\xi}\right).
\end{equation}
The Greek letter $\varepsilon$ depicts the Weyl scalar which has considerable significance in interpreting the tidal forces. Its value for our case reads as
\begin{align}\nonumber
\varepsilon&=-\frac{{\nu}''e^{-\lambda}}{4}-\frac{\nu'^2
e^{-\lambda}}{8}+\frac{\lambda'\nu'e^{-\lambda}}{8}+\frac{\nu'e^{-\lambda}}{4r}-
\frac{\lambda'e^{-\lambda}}{4r}-\frac{e^{-\lambda}}{2r^2}-\frac{1}{2r^2}.
\end{align}
Exploiting the equations of motion stated in Eqs. (\ref{2f})-(\ref{4f}), we arrive at the following outcome which connects the mass function with the
fluid variables, the Weyl scalar and an additional term describing the consequences of the electromagnetic field
\begin{equation}\label{8f}
\frac{3m}{r^3}=4\pi|\mu|+4\pi\Pi-\varepsilon-\frac{5\tilde{s}^2}{2r^4}
\end{equation}
Differentiating w.r.t. the radial component and utilizing the definition of mass function presented in Eq. (\ref{7f}), we acquire
\begin{equation}\label{8*f}
\varepsilon=4\pi\Pi+\frac{4\pi}{r^3}\int_0^r |\mu|'r^3 dr+\frac{3}{2r^3}\int_0^r\frac{\tilde{s}^2}{r^2}dr-\frac{\tilde{s}^2}{r^4}.
\end{equation}
Substituting this value for $\varepsilon$ in Eq. (\ref{8f}), we obtain
\begin{equation}\nonumber
m=\frac{4\pi |\mu|r^3}{3}-\frac{4\pi}{3}\int_0^r |\mu|'r^3 dr-\frac{1}{2}\int_0^r\frac{\tilde{s}^2}{r^2}dr-\frac{\tilde{s}^2}{2r}.
\end{equation}
The above mentioned expression resembles Eq. (44) in \cite{26} with a significant difference that it incorporates the effects of electromagnetic field
indicated by the last two factors.
The gravitational mass that acts as a source of gravitational field is the active gravitational mass and its expression for the case of charged
hyperbolically symmetric matter is given as
\begin{equation}\label{9f}
m_T=\int^{2\pi}_0 \int^\pi_0 \int^r_0 r^2 Sinh \theta e^\frac{\nu+\lambda}{2}(T^0_0+\tilde{S}^0_0-T^1_1-\tilde{S}^1_1-2T^2_2-2\tilde{S}^2_2)d\tilde{r}d\theta d\phi,
\end{equation}
where $T_0^0,~T^1_1,~T^2_2$ denote the usual stress-energy tensor constituents whereas $\tilde{S}^0_0,~\tilde{S}^1_1,~\tilde{S}^2_2$ denote the Einstein-Maxwell tensor components.
Solving Eq. (\ref{9f}), we acquire the following outcome
\begin{equation}\label{10f}
m_T=2\pi(cosh\pi-1)\int^r_0 e^\frac{\nu+\lambda}{2}\tilde{r}^2 \left(-|\mu|+P_r+2P_\bot+\frac{\tilde{s}^2}{4\pi r^4}\right)d\tilde{r}.
\end{equation}
Employing Eqs. (\ref{2f})-(\ref{4f}) and then integrating w.r.t. $r$, we attain
\begin{equation}\label{11f}
m_T=\frac{cosh\pi-1}{4}\nu'r^2 e^\frac{\nu-\lambda}{2},
\end{equation}
which in combination with Eq. (\ref{7*f}) produces
\begin{equation}\label{12f}
m_T=\frac{cosh\pi-1}{2}\left(4\pi P_r r^3-m-\frac{\tilde{s}^2}{r}\right)e^\frac{\nu+\lambda}{2}.
\end{equation}
From Eq. (\ref{10f}), it can be noticed that $m_T$ is a negative quantity. Also, from Eq. (\ref{12f}), it can be observed that $m_T$ would be negative if
$4\pi P_r r^3-\tilde{s}^2/r<m$. This would in turn depict the repulsive nature of the charged gravitational field in the considered spacetime model.
The four-acceleration symbolized by $a_\eta$ defined as $a_\eta=V_{\eta;\xi}V^\xi$ takes the following form
\begin{equation}\nonumber
a_\eta=a K_\eta,
\end{equation}
where, the scalar associated with it reads $a=\frac{\nu' e^\frac{-\lambda}{2}}{2}$. Employing this value in Eq. (\ref{11f}), we attain
\begin{equation}\nonumber
a=\frac{2m_T e^\frac{-\nu}{2}}{r^2 (cosh\pi-1)}.
\end{equation}
Since $m_T$ is negative for $4\pi P_r r^3-\tilde{s}^2/r<m$, it reveals that the four acceleration is also negative, i.e., it is directed inwards radially.
This demonstrates that the gravitational force has repulsive nature. Now, taking the radial derivative of Eq. (\ref{9f}) and utilizing Eq. (\ref{12f}),
we acquire
\begin{equation}\nonumber
m'_T-\frac{3m_T}{r}= -\left({\frac{cosh\pi-1}{2}}\right)r^2 e^\frac{\nu+\lambda}{2}\left(\varepsilon+4\pi\Pi-\frac{\tilde{s}^2}{4\pi r^4}\right).
\end{equation}
Integrating w.r.t. the radial component, we attain
\begin{equation}\nonumber
m_T=(m_T)_{\Sigma_e}\left(\frac{r^3}{r^3_{\Sigma^e}}\right)+\left(\frac{cosh\pi-1}{2}\right)r^3 \int^{r_{\Sigma^e}}_r
\frac{e^\frac{\nu+\lambda}{2}}{\tilde{r}}\left(\varepsilon+4\pi\Pi-\frac{\tilde{s}^2}{4\pi \tilde{r}^4}\right)d\tilde{r},
\end{equation}
which on substituting the value from Eq. (\ref{8*f}) reads
\begin{equation}\label{13f}
m_T=(m_T)_{\Sigma^e}\left(\frac{r^3}{r^3_{\Sigma^e}}\right)+\left(\frac{cosh\pi-1}{2}\right)r^3 \int^{r_{\Sigma^e}}_r \frac{e^\frac{\nu+\lambda}{2}}
{\tilde{r}}\left[\frac{4\pi}{\tilde{r}^3}
\int^r_0 |\mu|' \tilde{r}^3 d\tilde{r}+8\pi\Pi-\frac{\tilde{s}^2}{\tilde{r}^4}\right]d\tilde{r}
\end{equation}
Again, this result resembles Eq. (54) in \cite{26} with an obvious difference of charge terms in the last integral on right-hand-side of the expression.
This points towards the fact that the electromagnetic field has impact on the active gravitational mass of the configuration.

\section{The Orthogonal Division of Riemann Tensor and Structure Parameters}

As introduced by Bel \cite{28} and used in their respective analysis by \cite{15,16,18,19,20,29,30}, the technique of orthogonally dividing the Riemann tensor
into three constituent parts assists to attain three tensor quantities. These are beneficial in the sense that they can be utilized to acquire SS. Such functions, using the gravitational equations, are related with the physical characteristics of matter content to explicate the evolutionary phases as well as structure development of such configurations. The three tensors procured from such a division are given as under
\begin{eqnarray}\nonumber
&&Y_{\alpha\xi}=R_{\alpha\gamma\xi\delta}u^\gamma
u^\delta,\\\nonumber
&&Z_{\alpha\xi}=^*R_{\alpha\gamma\xi\delta}u^\gamma
u^\delta=\frac{1}{2}\eta_{\alpha\gamma\epsilon\rho}R^{\epsilon\rho}_{\xi\delta}u^\gamma
u^\delta,\\\nonumber
&&X_{\alpha\xi}=^*R^*_{\alpha\gamma\xi\delta}u^\gamma
u^\delta=\frac{1}{2}\eta_{\alpha\gamma}^{\epsilon\rho}R^*_{\epsilon\rho\xi\delta}u^\gamma
u^\delta,
\end{eqnarray}
with
\begin{equation}\nonumber
R^*_{\alpha\xi\gamma\delta}=\frac{1}{2}\eta_{\epsilon\omega\gamma\delta}R^{\epsilon\omega}_{\alpha\xi}.
\end{equation}
Utilizing the Einstein-Maxwell equations, we write
\begin{align}\nonumber
R^{\sigma\xi}_{\tau\beta}=C^{\sigma\xi}_{\tau\beta}+16\pi
(\left.T\right.^{[\sigma}_{[\tau}\delta^{\xi]}_{\beta]}+\left.\tilde{S}\right.^{[\sigma}_{[\tau}\delta^{\xi]}_{\beta]})+8\pi
\left.T\right.\left(\frac{1}{3}\delta^\sigma_{[\tau}\delta^\xi_{\beta]}-\delta^{[\sigma}_{[\tau}\delta^{\xi]}_{\beta]}
\right).
\end{align}
Making use of the stress-energy tensor, we may perform the required division as follows
\begin{equation}\nonumber
R^{\sigma\xi}_{\tau\beta}=R^{\sigma\xi}_{(I)\tau\beta}+R^{\sigma\xi}_{(II)\tau\beta}+R^{\sigma\xi}_{(III)\tau\beta},
\end{equation}
having the following values
\begin{align}\nonumber
R^{\sigma\xi}_{(I)\tau\beta}&=16\pi\left(\mu+\frac{\tilde{s}^2}{8\pi r^4}\right)V^{[\sigma}V_{[\tau}\delta^{\xi]}_{\beta]}
-16\pi \left(P+\frac{\tilde{s}^2}{8\pi r^4}\right)
h^{[\sigma}_{[\tau}\delta^{\xi]}_{\beta]}+8\pi (\mu-3P)\\\nonumber &\times\left(\frac{1}{3}\delta^\sigma_{[\tau}
\delta^\xi_{\beta]}-\delta^{[\sigma}_{[\tau}\delta^{\xi]}_{\beta]}\right),
\\\nonumber
R^{\sigma\xi}_{(II)\tau\beta}&=16\pi\left(\Pi-\frac{\tilde{s}^2}{4\pi r^4}\right)\left[K^{[\sigma} K_{[\tau}\delta^{\xi]}_{\beta]}
 +\frac{1}{3} h^{[\sigma}_{[\tau}\delta^{\xi]}_{\beta]}\right],
\\\nonumber
R^{\sigma\xi}_{(III)\tau\beta}&=4V^{[\sigma}V_{[\tau}E^{\xi]}_{\beta]}-\epsilon^{\sigma\xi}_\mu
\epsilon_{\tau\beta\nu}E^{\mu\nu}.
\end{align}
The explicit expressions for the three tensors read as
\begin{align}\label{14f}
Y_{\alpha\xi}&=E_{\alpha\xi}+4\pi\left(\Pi-\frac{\tilde{s}^2}{r^4}\right)\left(K_\alpha K_\xi+\frac{h_{\alpha\xi}}{3}\right)
+\frac{4\pi}{3}(\mu+3P)h_{\alpha\xi},\\\label{15f} X_{\alpha\xi}&=-\frac{8\pi|\mu|}{3}h_{\alpha\xi}
+\left(\Pi-\frac{\tilde{s}^2}{r^4}\right)\left(K_\alpha K_\xi+\frac{h_{\alpha\xi}}{3}\right)-E_{\alpha\xi},\\\label{16f}
Z_{\alpha\xi}&=0.
\end{align}
The tensors quantities mentioned in  Eqs. ({\ref{14f}), (\ref{15f}}) can further be divided into trace and trace-free constituents as
\begin{align}\nonumber
X_{\alpha\xi}&=\frac{h_{\alpha\xi}}{3}X_T +\left(K_\alpha K_\xi +\frac{h_{\alpha\xi}}{3}\right)X_{TF},\\\nonumber
Y_{\alpha\xi}&=\frac{h_{\alpha\xi}}{3}Y_T +\left(K_\alpha K_\xi +\frac{h_{\alpha\xi}}{3}\right)Y_{TF}.
\end{align}
The computation of trace and trace-free parts of both the quantities lead us to the following results
\begin{align}\label{17f}
X_T&=-8\pi|\mu|+\frac{\tilde{s}^2}{r^4},\\\label{18f}
X_{TF}&=4\pi\Pi-\varepsilon-\frac{\bar{s}^2}{r^4},\\\label{19f}
Y_T&=4\pi(-|\mu|+3P)+\frac{\tilde{s}^2}{r^4},\\\label{20f}
Y_{TF}&=4\pi\Pi+\varepsilon-\frac{\tilde{s}^2}{r^4}.
\end{align}
Equations (\ref{17f})-(\ref{20f}) resemble Eqs. (69,70) and (72,73) in \cite{26} with a difference that the influence of electromagnetic field
absent there can clearly be seen in the stellar equations of this work. Both the trace-free parts, using Eq. (\ref{8*f}) can also be written as
\begin{align}\nonumber
X_{TF}&=-\frac{4\pi}{r^3}\int^r_0 \tilde{r}^3|\mu|'d\tilde{r}-\frac{3}{2r^3}\int^r_0\frac{\tilde{s}^2}{\tilde{r}^2}d\tilde{r},\\\label{20*f}
Y_{TF}&=8\pi\Pi+\frac{4\pi}{r^3}\int^r_0 \tilde{r}^3|\mu|'d\tilde{r}+\frac{3}{2r^3}\int^r_0\frac{\tilde{s}^2}{\tilde{r}^4}d\tilde{r}-\frac{2\tilde{s}^2}{r^4}.
\end{align}
From Eqs. (\ref{18f}) and (\ref{20f}), we acquire
\begin{equation}\nonumber
X_{TF}+Y_{TF}=8\pi\Pi-\frac{\tilde{2s}^2}{r^4}.
\end{equation}
To interpret the physical sense of $Y_T,~Y_{TF}$, we make use of Eq. (\ref{13f}) and obtain
\begin{align}\nonumber
m_T&=(m_T)_{\Sigma e} \left(\frac{r}{r_{\Sigma e}}\right)^3+\left(\frac{cosh\pi-1}{2}\right)r^3\int^{r_{\Sigma e}}_r
\frac{e^\frac{\nu+\lambda}{2}}{\tilde{r}}\left(Y_{TF}+\frac{\tilde{s}^2}{\tilde{r}^4}\right)d\tilde{r},\\\nonumber
m_T&=\frac{cosh\pi-1}{2}\int_0^r \tilde{r}^2 e^{(\nu+\lambda)/2}Y_Tdr.
\end{align}
These results exhibit that $Y_{TF}$ governs the impact of density inhomogeneity and stress anisotropy on the active gravitational mass of the matter content. Also, $Y_T$ is found to be proportional to active gravitational mass density.

\section{Static Hyperbolically Symmetric Solutions}

This section encompasses a general technique that may be utilized to indicate any kind of hyperbolically symmetric static solution with the help of two
generating functions. From the last two equations of motion, i.e., Eqs. (\ref{3f}) and (\ref{4f}), we acquire
\begin{equation}\nonumber
8\pi(P_r-P_\bot)-\frac{2\tilde{s}^2}{r^4}=\frac{1+e^{-\lambda}}{r^2}-\frac{e^{-\lambda}}{2}\left(\nu''+\frac{\nu'^2}{2}-\frac{\lambda'\nu'}{2}
-\frac{\nu'}{r}-\frac{\lambda'}{r}\right).
\end{equation}
As mentioned in \cite{26}, we make use of new functions $\frac{\nu'}{2}=z-\frac{1}{r}$ and $y=e^{-\lambda}$ to transform the above equation into the
following differential equation in $y$ as
\begin{align}\nonumber
y'+y\left[\frac{4}{r^2z}+2z+\frac{2z'}{z}-\frac{6}{r}\right]=\frac{2}{z}\left[\frac{1}{r^2}-8\pi\Pi+\frac{2\tilde{s}^2}{r^4}\right],
\end{align}
whose integration produces
\begin{align}\label{21f}
e^{\lambda(r)}=\frac{z^2
e^{\int\left(2z+\frac{4}{zr^2}\right)dr}}{r^6\left[2\int\left\{ z\left(\frac{1-8\pi\Pi r^2+2\tilde{s}/r^2}{r^8}\right)
e^{\int\left(2z+\frac{4}{zr^2}\right)dr}\right\}dr+\bar{C}_1\right]}.
\end{align}
It can be noticed from Eq. (\ref{21f}) that the functions $\Pi$ and $z$ are the generating functions which can be employed to completely describe any kind
of static solution for our case. The physical fluid variables can now be written down as
\begin{align}\nonumber
4\pi|\mu|&=\frac{m'}{r^2}+\frac{\tilde{s}\tilde{s}'}{r^3},\\\nonumber
4\pi P_r&=\frac{zr(2m-r+\tilde{s}^2/r)-m+r}{r^3},\\\nonumber
8\pi P_\bot&=\left(\frac{2mr-r^2+\tilde{s}^2}{r^2}\right)\left[z'+\frac{1}{r^2}+z^2-\frac{z}{r}\right]+z\left[\frac{m'}{r}-\frac{m}{r^2}+
\frac{\tilde{s}\tilde{s}'}{r^2}-\frac{\tilde{s}^2}{r^3}\right]-\frac{\tilde{s}^2}{r^4}.
\end{align}

\subsection{The Conformally Flat Solution}

From Eq. (\ref{8*f}), it can be noticed that the Weyl scalar has considerable influence on the structure of matter content. So, we may take into account
a particular case, i.e., $\varepsilon=0$. By utilizing the definition of the Weyl scalar mentioned before Eq. (\ref{8f}), we may write
\begin{equation}\nonumber
\frac{\partial}{\partial r}\left\{\frac{\nu' e^{{-\lambda}}}{2r}\right\}+e^{-\nu-\lambda} \frac{\partial}{\partial r}\left\{\frac{\nu' e^\nu}{2r}\right\}
-\frac{\partial}{\partial r}\left\{\frac{1+e^{-\lambda}}{r^2}\right\}=0.
\end{equation}
The variables $y=e^{-\lambda}$ and $\frac{\nu'}{2}=\frac{w'}{w}$ mentioned previously in \cite{26} are utilized to render the following outcome
\begin{equation}\nonumber
y'+2\left(\frac{w''-\frac{w'}{r}+\frac{w}{r^2}}{w'-\frac{w}{r}}\right)y+\frac{2w}
{\left(w'-\frac{w}{r}\right)r^2}=0,
\end{equation}
which can be integrated to obtain the formal solution for $y$ as
\begin{equation}\label{22f}
y=e^{-\int f_1(r)dr}\left(\int e^{-\int f_1(r)dr}f_2(r)dr+\bar{C}_2\right),
\end{equation}
here, $\bar{C}_2$ signifies the integration constant and the values for $f_1(r),~f_2(r)$ are given as
\begin{eqnarray}\nonumber
&&f_1(r)=2 \frac{d}{dr}\left[ln\left(w'-\frac{w}{r}\right)\right],\\\nonumber
&&f_2(r)=\frac{-2w} {\left(w'-\frac{w}{r}\right)r^2}.
\end{eqnarray}
Transforming Eq. (\ref{22f}) by utilizing original variables, we attain
\begin{equation}\label{23f}
\frac{\nu'}{2}-\frac{1}{r}=\frac{e^{\lambda/2}}{r}\sqrt{-1+r^2\alpha_1 e^{-\nu}}.
\end{equation}
Here, $\alpha_1$ indicates the constant of integration. The junction conditions can be used to determine its value as follows
\begin{equation}\nonumber
\alpha_1=\frac{(3M-r_{\Sigma^e}+2Q^2/r_{\Sigma^e})^2+r_{\Sigma_e} (2M-r_{\Sigma^e}+Q^2/r_{\Sigma^e})}{r^4_{\Sigma^e}}.
\end{equation}
The integration of Eq. (\ref{23f}) produces
\begin{equation}\nonumber
e^\nu=\alpha_1 r^2 Sin^2\left(\int\frac{e^{\lambda/2}}{r}dr+\zeta\right),
\end{equation}
with $\zeta$ being the constant of integration. Again junction conditions can be employed to obtain
\begin{align}\nonumber
\zeta&=Sin^{-1}\left[r_{\Sigma^e}\left\{\frac{\frac{2M}{r_{\Sigma^e}}-1+
\frac{Q^2}{r_{\Sigma^e}^2}}{(3M-r_{\Sigma_e}+2Q^2/r_{\Sigma^e})^2+r_{\Sigma^e} (2M-r_{\Sigma^e}+Q^2/r_{\Sigma^e})}
\right\}^{1/2}\right]\\\nonumber &-\left[\int\frac{e^{\lambda/2}}{r}dr\right]_{\Sigma^e}.
\end{align}
The conformally flat condition provides only one generating function and in order to determine a particular model, we must consider an additional extreme condition,
i.e., $P_r=0$. Utilizing this assumption along with Eq. (\ref{3f}), we yield
\begin{equation}\label{24f}
\nu'=-\frac{(1+e^\lambda+\tilde{s}^2/r^2)}{r}.
\end{equation}
Making use of the values for $\nu'$ and $\varepsilon$, the following outcome is worked out
\begin{align}\nonumber
\left(1+e^\lambda+\frac{\tilde{s}^2}{r^2}\right)^2+8(1+e^\lambda)+3\lambda'r-\lambda're^\lambda-\frac{4\tilde{s}\tilde{s}'}{r}
-\frac{8\tilde{s}^2}{r^2}+\frac{\lambda'\tilde{s}^2}{r}=0.
\end{align}
Putting $e^{-\lambda}=2g-1$, the above result can be re-written as
\begin{equation}\label{25f}
g(9g-4)-g'r(3g-2)+G_1(r)=0,
\end{equation}
where
\begin{align}\nonumber
G_1(r)&=g^2\left(\frac{4\tilde{s}^4}{r^4}-\frac{8\tilde{s}^2}{r^2}-\frac{16\tilde{s}\tilde{s}'}{r}\right)
+g\left(-\frac{4\tilde{s}^4}{r^4}+\frac{12\tilde{s}^2}{r^2}+\frac{16\tilde{s}\tilde{s}'}{r}\right)+\left(\frac{\tilde{s}^4}{r^4}
-\frac{4\tilde{s}\tilde{s}'}{r}-\frac{8\tilde{s}^2}{r^2}\right)\\\nonumber &-g'\left(\frac{2\tilde{s}^2}{r}\right).
\end{align}
Eqs. (\ref{25f}) can be integrated to produce
\begin{equation}\nonumber
\bar{C}_3r^6=\frac{4g^3}{(9g-4)}+r^6\int H_1(r)dr,
\end{equation}
with $\bar{C}_3$ being the constant of integration and $H_1(r)=24r^5g^2G_1(r)$. Utilizing Eqs. (\ref{23f}) and (\ref{24f}), we acquire
\begin{equation}\nonumber
e^\nu=\frac{\alpha_1r^2(2g-1)}{9g^2-4g+\frac{\tilde{s}^2}{4r^2}(2g-1)^2-\frac{(1-3g)(2g-1)\tilde{s}^2}{r^2}}.
\end{equation}
The physical fluid variables for this case read
\begin{align}\nonumber
|\mu|&=\frac{6g^2-3g+(G_1(r)/2)}{2\pi r^2(3g-2)}-\frac{\tilde{s}^2}{2r^4},\\\nonumber
P_\bot&=\frac{3g^2+(gr/2)G_1(r)}{4\pi r^2(3g-2)}+A_1,
\end{align}
with the value of $A_1$ being equal to
\begin{equation}\nonumber
A_1=\frac{\tilde{s}^2(9g^2-4g+G_1(r))}{r^4(3g-2)}-\frac{\tilde{s}^4}{r^4}+\frac{2g-1}{2}\left\{\frac{3\tilde{s}^2}{r^4}
-\frac{2\tilde{s}\tilde{s}'}{r^3}+\frac{3\tilde{s}^4}{2r^6}+\frac{2g\tilde{s}^2}{r^4(2g-1)}-\frac{\tilde{s}^2}{r^3}\right\}.
\end{equation}
Finally, the generating functions for such a model are
\begin{eqnarray}\nonumber
&&z(r)=\frac{g-1}{r(2g-1)}-\frac{\tilde{s}^2}{r^3},\\\nonumber
&&\Pi(r)=-\left\{\frac{3g^2+(gr/2)G_1(r)}{4\pi r^3(3g-2)}+A_1\right\}.
\end{eqnarray}

\subsection{A model with Zero Complexity Factor}

To investigate the static homogeneous sources, Herrera \cite{35} utilized a factor called the complexity factor responsible for determining the complexity of matter. For our case, the function $Y_{TF}$ administers the complexity of matter source. This section demonstrates another interesting solution by taking into account the zero $Y_{TF}$ parameter, i.e., zero complexity factor of the configuration. In order to acquire a particular model, we need to assume another restriction besides this. So, we let $P_r=0$ so that Eq. (\ref{3f}) implies
\begin{equation}\label{26f}
\nu'=\frac{-2g}{r(2g-1)}-\frac{\tilde{s}^2}{r^3},
\end{equation}
with the value of $g$, same as the one defined for the case of conformally flat solutions. Employing the condition $Y_{TF}=0$, we acquire
\begin{equation}\nonumber
m_T=(m_T)_{\Sigma_e}\left(\frac{r}{r_{\Sigma^e}}\right)^3+\left(\frac{cosh \pi-1}{2}\right)r^3\int_r^{r_{\Sigma^e}}
\frac{e^{(\nu+\lambda)/2}\tilde{s}^2}{\tilde{r}^5}d\tilde{r}.
\end{equation}
Utilizing Eqs. (\ref{11f}) and (\ref{26f}), the value for the metric coefficient $e^\nu$ becomes
\begin{align}\nonumber
e^\nu&=\frac{1}{(cosh\pi -1)^2}\left[\frac{4g^2r^2}{}(2g-1)+\frac{\tilde{s}^4(2g-1)}{r^3}+2g\tilde{s}^2\right]^{-1}
\left[(m_T)_{\Sigma^e}\left(\frac{r}{r_{\Sigma^e}}\right)^3+\right.\\\nonumber &\;\;\;\left.\left(\frac{cosh \pi-1}{2}\right)r^3\int_r^{r_{\Sigma^e}}
\frac{e^{(\nu+\lambda)/2}\tilde{s}^2}{\tilde{r}^5}d\tilde{r}\right]^2.
\end{align}
The restriction $Y_{TF}=0$ also produces
\begin{equation}\nonumber
-g(5g-2)+g'r(-1+g)+A_2,
\end{equation}
where
\begin{equation}\nonumber
A_2=(2g-1)\left[\frac{\tilde{s}\tilde{s}'}{r^3}-\frac{3\tilde{s}^2}{2r^4}+\frac{\tilde{s}^4}{4r^6}-\frac{g\tilde{s}^2}{2r^4}+\frac{\tilde{s}^2}{r^3}
\right]+\frac{\tilde{s}^2}{r^4}.
\end{equation}
We can integrate this equation to produce
\begin{equation}\nonumber
\bar{C}_4r^{10}=\frac{g^5}{(5g-2)^3}+r^{10}\int H_2(r)dr,
\end{equation}
with $\bar{C}_4$ indicating the integration constant and $H_2(r)=10r^9g^4G_2(r)$. The fluid variables for such a model may be expressed as
\begin{align}\nonumber
|\mu|&=\frac{3g(2g-1)-G_2(r)}{4\pi r^2(g-1)},\\\nonumber
P_\bot&=\frac{1}{8\pi r^2(g-1)}\left(3g^2+\frac{grG_2(r)}{1-2g}\right)+A_3,
\end{align}
with $A_3$ being equal to
\begin{equation}\nonumber
A_3=\frac{\tilde{s}^2}{r^3}\left(\frac{-g(5g-2)+G_2(r)}{r(1-g)}\right)+\frac{2g\tilde{s}^2}{r^4(2g-1)}-\frac{2\tilde{s}^2}{r^4}
-\frac{2\tilde{s}\tilde{s}'}{r^3}+\frac{\tilde{s}^4}{2r^6}.
\end{equation}
Furthermore, the generating functions for such a kind of model take the following form
\begin{align}\nonumber
z&=\frac{g-1}{r(2g-1)}-\frac{\tilde{s}^2}{2r^3},\\\nonumber
\Pi&=-\frac{1}{8\pi r^2(g-1)}\left(3g^2+\frac{grG_2(r)}{1-2g}\right)-A_3.
\end{align}

\subsection{The Stiff Equation of State}

An equation of state in which a slight change in the density of the system causes a rapid increase in its pressure is termed as stiff equation of state. This section incorporates such an equation of state to produce certain solutions that were first proposed by Zeldovich \cite{34}. Such a solution presumes that amount of stress equals the energy density of the system, i.e., $|\mu|=P_r$. Under such considerations, Eq. (\ref{7**f}) reads
\begin{equation}\label{27f}
P_r'+\frac{2\Pi}{r}-\frac{\tilde{s}\tilde{s}'}{4\pi r^4}=0.
\end{equation}
To work out a particular solution, we need additional restrictions. For this purpose, we first assume $P_\bot=0$ and later, we would take $Y_{TF}=0$.
\begin{itemize}
\item{\textbf{When $P_\bot=0$}}

Under this restriction, the integration of Eq. (\ref{27f}) provides
\begin{align}\nonumber
P_r=\frac{K}{r^2}+\frac{1}{r^2}\int_0^r \frac{\tilde{s}\tilde{s}'}{4\pi r^2}dr \quad\quad
\Rightarrow|\mu|=\frac{K}{r^2}+\frac{1}{r^2}\int_0^r \frac{\tilde{s}\tilde{s}'}{4\pi r^2}dr,
\end{align}
with $K$ denoting the integration constant. Utilizing Eqs. (\ref{6f}), (\ref{7f}) and (\ref{7*f}), we attain
\begin{align}\nonumber
m=4\pi Kr-\int_0^r\frac{\tilde{s}\tilde{s}'}{r}dr; \quad\quad e^{-\lambda}=8\pi K-1-\frac{2}{r}\int_0^r \frac{\tilde{s}\tilde{s}'}{r}dr-
\frac{\tilde{s}^2}{r^2}.
\end{align}
The generating functions become
\begin{equation}\nonumber
z=\frac{1}{r}\left[1-\frac{1}{2m-r+\tilde{s}^2/r}\left(\int_0^r \frac{\tilde{s}\tilde{s}'}{r}dr+\frac{\tilde{s}^2}{r^2}\right)\right]; \quad\quad \Pi=\frac{K}{r^2}.
\end{equation}
\item{\textbf{When $Y_{TF}=0$}}\\

In addition to the stiff equation of state, we now assume the zero complexity factor condition. Utilizing this condition in Eq. (\ref{20*f}), and substituting the result in Eq. (\ref{27f}), we attain
\begin{equation}\nonumber
P_r''+\frac{3P_r'}{r}+S_1=0,
\end{equation}
with
\begin{equation}\nonumber
S_1= -\frac{7\tilde{s}^2}{8\pi r^6}+\frac{\tilde{s} \tilde{s}'}{\pi r^5}-\frac{\tilde{s} \tilde{s}''}{4\pi r^4}-\frac{\tilde{s}'^2}{4\pi r^4}=0.
\end{equation}
The solution of this equation becomes equal to
\begin{equation}\label{28f}
P_r=\frac{b}{r^2}-a+S^\star.
\end{equation}
The alphabets $a$ and $b$ indicate positive integration constants. From the definition of mass function, we get
\begin{equation}\nonumber
S^\star=-a-\int\frac{1}{r^3}\left(\int S_1 r^3 dr\right)dr.
\end{equation}
Eq. (\ref{28f}) can be utilized to obtain the value for variable $\lambda$ and Substituting its value in Eq. (\ref{7*f}), the value for $\nu$ may be acquired.\\
\begin{equation}\nonumber
m=4\pi r\left(b-\frac{ar^2}{3}\right)+4\pi\int^r_0 \left(S^\star r^2+\frac{\tilde{s} \tilde{s}'}{4\pi r}\right)dr.
\end{equation}
Further, if we assume that the matter content is bounded from the exterior region by the boundary $\Sigma^e$, then we get
\begin{equation}\nonumber
P_r=b\left[\frac{1}{r^2}-\frac{1}{r^2_{\Sigma^e}}\right],
\end{equation}
along with
\begin{equation}\nonumber
m=\frac{4\pi br}{3 r^2_{\Sigma^e}}(3r^2_{\Sigma^e}-r^2)+\int^r_0 \frac{\tilde{s} \tilde{s}'}{r}dr.
\end{equation}
We can also obtain an interesting relationship between the stress component $P_r$ and radius $r_{\Sigma^e}$ along with additional terms indicating the consequences of electromagnetic field.
\begin{equation}\nonumber
4\pi P_r r^3-m-\frac{\tilde{s}^2}{r}=\frac{-8\pi br^3}{3r^2_{\Sigma e}}-\int^r_0 \frac{\tilde{s} \tilde{s}'}{r}dr-\frac{\tilde{s}^2}{r}.
\end{equation}
The tangential pressure $P_\bot$ for such a solution reads.
\begin{equation}\nonumber
P_\perp= -\frac{b}{r^2_{\Sigma e}}-\frac{\tilde{s} \tilde{s}'}{4\pi r^3}.
\end{equation}
\end{itemize}

\section{Conclusion}

We contemplate a hyperbolically symmetric spacetime occupied with static matter source anisotropic in pressure and bounded from the exterior by a hypersurface $\Sigma^e$ whose
equation can be mathematically expressed as $r=r_{\Sigma^e}=\textmd{constant}$. The middle region is not occupied by matter which
points towards the fact that a cavity is present there. This implies that the matter under consideration is also bounded from the interior, the mathematical
equations of which can be written as $r=r_{\Sigma^i}=\textmd{constant}$. Choosing the comoving coordinates for our work, we utilize the orthonormal tetrad and write the
locally Minkowski coordinates to express the constituents of stress-energy tensor. Since the matter source is bounded both from interior and exterior, the
satisfaction of Darmois conditions \cite{25} is necessary. From the exterior region, the Schwarzschild metric mentioned before section \textbf{3} is taken into account and for interior region, the Minkowski spacetime is used. We specify the gravitational equations influenced by the electromagnetic field and utilized the constituents of stress-energy tensor and Einstein-Maxwell tensor to attain hydrostatic equilibrium equation. The distribution of mass inside the matter content indicated by $m(r)$ is evaluated, the expression of which demonstrates that the energy density must necessarily be negative. The curvature tensors, i.e., the Weyl and the Riemann tensor are obtained. The Weyl scalar associated with the description of tidal forces experienced by the object is linked mathematically with the energy density inhomogeneity and stress anisotropy.

The active gravitational mass for an arbitrary component of our static matter content inside the boundary $\Sigma^e$ is manipulated. Four SS $X_T,~X_{TF},~Y_T,~Y_{TF}$, extracted from the division of Riemann tensor, are found to be connected with the fundamentals of fluid. Finally, a general formalism is
constructed to demonstrate static solutions of considered hyperbolically symmetric source with the help of two generating functions. Such solutions include the conformally
flat solutions, the models with zero complexity factor and the solutions acquired by considering the stiff equation of state. The middle region is excluded during all of the above analysis.

To cut a long story short, we conducted a comprehensive assessment on the physical characteristics of static matter source possessing hyperbolical symmetry. From the manipulations, we come forth with the fact that such kind of matter source may be endowed with stress anisotropy (only two unequal stresses) and negative energy density. It is an already known fact that the hyperbolically symmetric matter source fails to occupy the space inside the horizon which results in the formation of a vacuole or a cavity in the middle region independent of whether the source has regular or irregular energy density. So, either the middle region must be included in the cavity or it should be defined by using another kind of matter source. From the point of view of classical physics, the energy density
of a matter source is supposed to be positive only. However, in some astrophysical scenarios, negative value of energy density may be involved, e.g., in the
interior region of a Gravastar. From Eq. (\ref{7f}), it can be noted that the energy density is necessarily negative which is the clear-cut violation of energy
condition (weak). From Eqs. (\ref{11f}) and (\ref{12f}), we infer that Tolman mass is negative, i.e., if $4\pi P_r r^3 <m$.

\end{document}